# arXiv preprint submitted for LAK20 conference: Adaptive Learning Guidance System (ALGS)

Ghada El-Hadad, Doaa Shawky, and Ashraf Badawi,

Gfawzy, dshawky, and abadawi@zewailcity.edu.eg

Center of Learning Technologies, Zewail City of Science and Technology

**Abstract**

This poster presents the conceptual framework of the Adaptive Learning Guidance System ALGS. The system aims to propose a model for adaptive learning environments where two major concerns arising from past studies are being addressed; the marginal role of the teacher, and the need for a big data approach. Most past studies marginalized the teacher's role in adaptive learning system, particularly the online ones. The most notable quality about ALGS is empowering the teacher with the capability of having input in all stages. This is where the hybrid recommendation system plays a crucial role in the 3-stage ALGS architecture. The second issue addressed is the need for big data to enhance the system functionality. The more the data collected by the system, the more efficient its adaptation functionality which makes it difficult for a first-time-run system and/or a first-time user. Accordingly, collaborative filtering is used at first until adequate data about the user interaction are collected. ALGS architecture consists of a user, content, and 3-stage adaptation models.

*adaptive learning; tutoring system; machine learning; online learning environment; adaptation model;*

1.  **ALGS architecture**

The conceptual architecture of ALGS is based on: User, Content, and a 3-stage adaptation models. The user model represents the data about the learner stored in the system (what the system adapts to) (Aleven, McLaughlin, Glenn, & Koedinger, 2016). The content model is the course content offered by the system and its hierarchical structure and logical order (what the system adapts) (Aleven et al., 2016). The adaptation model sets the adaptation strategies and rules (how the system adapts) (Fakeeh, 2017). The adaptation model consists of 3 stages. The first stage is the collaborative filtering. The second stage is the hybrid recommendation system where machine and teacher recommendations are generated. The third stage is the personalization engine. The data coming from all 3 models are combined to optimize suggestions and predictions to both the learner and the teacher simultaneously. The Adaptive Learning Guidance System architecture is illustrated in Figure 1.

**2. ALGS architecture components**

*2.1. User Model*

Data about the user is stored to be retrieved later by the system to enhance the personalization procedure (Rodríguez, valle, & Duque, 2015). User attributes are Cognitive traits, learning preferences (Aleven et al., 2016), Knowledge level, Personal data (Rodríguez et al., 2015) as illustrated in Figure 2.

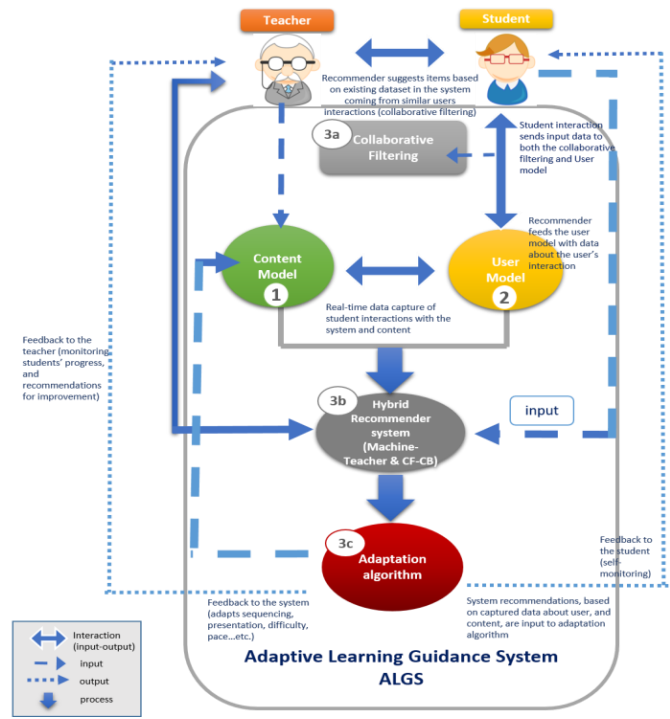

**Figure 1. Adaptive Learning Guidance System Architecture**

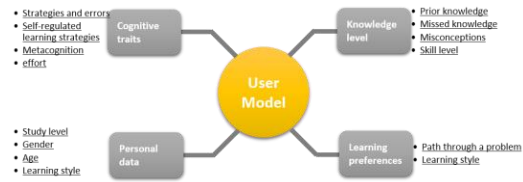

**Figure 2. User Model (Learner Attributes)**

## 2.2. Content Model

The model stores data about the knowledge to be presented to the learners and how it is organized in an adaptive manner to achieve the educational goals (Esichaikul, Lamnoi, & Bechter, 2011). The content model consists of content (Rodríguez et al., 2015), and delivery (Esichaikul et al., 2011).

*2.3. ALGS Adaptation Model*

*2.3.1. Stage 1: Collaborative Filtering (CF)*

When the hypothesized ALGS first runs, the user registers a new user profile. Collaborative filtering (CF) compares these data to a previously installed dataset of similar users and generates related suggestions for the user (Rodríguez et al., 2015). The dual function of CF is eliminating cold starts for first-time users on the one hand (Rodríguez et al., 2015), and allowing for more data gathering about users hence better adaptation on the other (Murray & Perez, 2015).

*2.3.2. Stage 2: Hybrid Recommendation System*

The hybrid recommendation system is hybrid on two levels: collaborative filtering (CF)-content-based (CB) hybrid recommender, and machine-teacher hybrid. The combined CF and CB recommenders together generate more effective predictions (Rodríguez et al., 2015). The result recommendations are based on user's interaction with the system, user's attributes, and data about similar users (Rodríguez et al., 2015). The recommendations address skills that the learner is yet to improve or likely to be weak at for further practice (Drachsler, Hummel, & Koper, 2007). In the machine-teacher hybridization, the generated recommendations are then delivered to the teacher to make decisions. The aroused issue with this step is the level of confidence in the recommendation provided by the system. In other words, the system may provide suggestions that the teacher does not understand the rationale behind in order to make an informed decision (Baker, 2019). Consequently, the recommender sends the teacher a combination of the user's input with the AI analytics. The teacher then decides how the recommendations are handled.

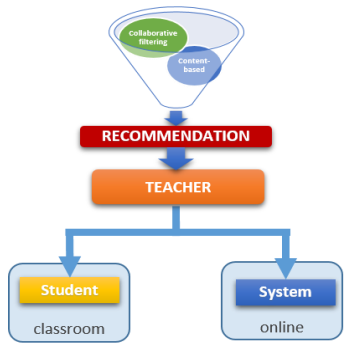

**Figure 3. Hybrid Recommendation System**

### *2.3.3. Stage 3: Personalization Engine*

The user attributes, content attributes, along with the 3 stages of adaptation are then integrated together to formulate efficient recommendations for both the student and the teacher about what material needs to be studied next (Fakeeh, 2017). The teacher has a crucial role at directing this stage being the decision maker of selecting, adding, or removing the next adaptation procedure.